# From Needs Assessment to Legal-Digital Empowerment: A Community-Based Framework for Indigenous Enterprises in Higher Education Extension Programs

Marilou O. Espina [a], Sales G. Aribe Jr. [b] ✉, Mark Ian M. Mukara [c], Joseph L. Abella [d], Nurwahyu Alamsyah [e]

[a] Bukidnon State University, Philippines
[b] Bukidnon State University, Philippines
[c] Bukidnon State University, Philippines
[d] Bukidnon State University, Philippines
[e] Universitas Muhammadiyah Yokyakarta, Indonesia

✉ Corresponding email: sg.aribe@buksu.edu.ph

## Abstract

Indigenous Peoples' Organizations (IPOs) contribute significantly to local economies, cultural preservation, and sustainable livelihoods within their ancestral communities. However, many IPOs remain marginalized in digital economies due to limited digital competencies, inadequate access to technology, and insufficient legal awareness related to online enterprise participation. This study examined the marketing practices, digital competencies, and legal-literacy needs of IPOs in the Philippines and developed a legal-digital empowerment framework for higher education extension and community engagement initiatives. Using a sequential mixed-methods design, data were collected through surveys and focus group discussions involving IPO members engaged in community-based enterprises. The study assessed existing marketing practices, identified competency and legal-awareness gaps, and determined capacity-building needs related to digital enterprise participation. Findings revealed that IPOs continue to rely heavily on traditional marketing channels, with limited use of digital platforms, inconsistent branding practices, and minimal engagement in strategic digital marketing activities. Across domains such as content creation, digital platforms, branding and packaging, marketing strategy, and legal awareness, most respondents demonstrated beginner-level proficiency. Participants also expressed concerns regarding intellectual property protection, online participation, and limited understanding of legal mechanisms related to digital enterprise engagement. Despite these limitations, IPO members showed strong willingness to participate in training and mentorship initiatives. The study proposes the Legal-Digital Empowerment Framework for Indigenous Community Enterprises (LDEF-ICE), which integrates digital competencies, legal literacy, institutional support systems, and rights-based market participation into a structured and context-responsive model for higher education extension programs. The study's novelty lies in reconceptualizing needs assessment as a foundation for legal-digital empowerment and framework development rather than as a purely descriptive process. Overall, the framework offers a scalable approach for promoting digital inclusion, legal empowerment, sustainable enterprise participation, and community resilience among indigenous populations.

# Introduction

IPOs play a vital role in sustaining local economies, preserving cultural heritage, and promoting environmentally responsible livelihoods within their ancestral domains. In the Bukidnon–Misamis Oriental landscape in the Philippines, IPOs engage in diverse community-based enterprises such as citronella oil production, abaca handicrafts, bamboo products, traditional weaving, and many other indigenous products. These initiatives reflect the integration of indigenous knowledge systems with sustainable livelihood practices, contributing not only to economic resilience but also to biodiversity conservation and cultural continuity.[1] Beyond their economic contribution, these enterprises also represent expressions of indigenous identity, self-determination, and community autonomy that require social, institutional, and legal protection within contemporary development contexts.[2]

Despite these contributions, IPOs continue to face persistent challenges in accessing broader markets and enhancing the competitiveness of their products. In contemporary market environments increasingly shaped by digital technologies, effective participation requires competencies in digital marketing, including content creation, branding, social media engagement, and data-driven promotion strategies. Digital marketing has become a critical driver of enterprise growth, particularly for micro, small, and community-based enterprises, as it enables wider market reach, customer engagement, and value creation.[3] However, marginalized and indigenous communities often remain at the periphery of digital transformation due to limited access to technology, insufficient digital literacy, and lack of structured learning opportunities.[4] These constraints

---

1 Manoj Kumar, Sweta Nisha Phukon, and Hukum Singh, "The Role of Communities in Sustainable Land and Forest Management," in *Forest Resources Resilience and Conflicts* (Elsevier, 2021), 305–18, https://doi.org/10.1016/B978-0-12-822931-6.00024-1; Lanhui Wang et al., "Transforming Forest Management through Rewilding: Enhancing Biodiversity, Resilience, and Biosphere Sustainability under Global Change," *One Earth* 8, no. 3 (March 2025): 101195, https://doi.org/10.1016/j.oneear.2025.101195.

2 United Nations, "United Nations Declaration on the Rights of Indigenous Peoples. Resolution 61/295," *United Nations General Assembly*, no. March (2007).

3 Elyria Kemp et al., "The Impact of Storytelling in Creating Firm and Customer Connections in Online Environments," *Journal of Research in Interactive Marketing* 15, no. 1 (May 2021): 104–24, https://doi.org/10.1108/JRIM-06-2020-0136; Budi Sunarso and Fahrina Mustafa, "Analysing the Role of Visual Content in Increasing Attraction and Conversion in MSME Digital Marketing," *Journal of Contemporary Administration and Management (ADMAN)* 1, no. 3 (November 2023): 193–200, https://doi.org/10.61100/adman.v1i3.79.

4 Sajda Qureshi, "Digital Transformation at the Margins: A Battle for the Soul of Self-Sovereignty," *Information Technology for Development* 28, no. 2 (April 2022): 215–29, https://doi.org/10.1080/02681102.2022.2062291; Bruce F. Richards et al., "Assessing the Needs of hinder their ability to leverage digital platforms for enterprise development and market expansion.

Beyond technological barriers, indigenous enterprises also encounter legal and regulatory challenges that affect their participation in digital economies. These include limited awareness of intellectual property rights, product labelling standards, consumer protection regulations, e-commerce policies, digital platform governance, and legal mechanisms that safeguard indigenous cultural products and identities from exploitation and misappropriation.[5] Indigenous entrepreneurs engaging in digital markets may also face vulnerabilities associated with unfair trade practices, online fraud, misuse of indigenous cultural representations, and unequal participation in digital commerce systems. Consequently, digital exclusion among indigenous communities should not be viewed solely as a technological issue but also as a legal and social justice concern linked to access, protection, and equitable participation.[6]

Within the Philippine context, these concerns are closely related to broader legal frameworks that recognize and protect indigenous rights, including the Indigenous Peoples' Rights Act of 1997 and policies implemented by the National Commission on Indigenous Peoples. These frameworks emphasize the protection of ancestral domains, cultural integrity, self-governance, and sustainable community development. However, while legal protections exist at the policy level, many indigenous enterprises continue to face difficulties translating these rights into practical economic participation within digitally mediated markets. This highlights the need for community-based legal education and empowerment

Utah Dairy Farmers to Inform Extension Programming," *Outcomes and Impact Quarterly* 3, no. 2 (July 2023), https://doi.org/10.59620/2995-2220.1059; Eva Banowati et al., "Digital Technology Use and Perceived Intergenerational Empowerment in Sustainable Rural Tourism: Evidence from a Youth-Dominated Sample in Mount Muria, Indonesia," *International Journal of Sustainable Development and Planning* 21, no. 8 (August 31, 2026): 3779–88, https://doi.org/10.18280/ijsdp.210827; Rofi Wahanisa et al., "Achieving Public Information Transparency in The Dissemination of Local Regulations," *Yuridika* 38, no. 3 (2023), https://doi.org/10.20473/ydk.v38i3.48345.

5 Lynette J. Chua and David M. Engel, "State and Personhood in Southeast Asia: The Promise and Potential for Law and Society Research," *Asian Journal of Law and Society* 2, no. 2 (November 2015): 211–28, https://doi.org/10.1017/als.2015.10.

6 Stephen Golub, *Legal Empowerment: Practitioners' Perspectives*, 2010; United Nations, "United Nations Declaration on the Rights of Indigenous Peoples. Resolution 61/295."

initiatives that strengthen both digital competencies and legal awareness among IPO members.

Higher education institutions (HEIs), particularly through extension and community engagement programs, are uniquely positioned to address these gaps by facilitating context-sensitive capacity-building initiatives. In this regard, extension programs are no longer limited to service delivery but are increasingly recognized as pedagogical platforms that integrate experiential learning, community-based education, legal empowerment, and knowledge co-creation.[7] Such approaches align with constructivist learning perspectives, where learners develop competencies through active engagement, contextual problem-solving, and collaboration with stakeholders.[8] Recent studies emphasize that HEIs play a critical role in bridging digital divides and promoting inclusive development by embedding real-world, community-based learning experiences into curricula, thereby enhancing both student learning outcomes and community empowerment.[9]

Moreover, contemporary perspectives on legal-community engagement emphasize that community development initiatives should integrate legal literacy and rights-based participation into educational and extension programs, particularly for marginalized communities.[10] In this context, legal empowerment refers to the process through which communities gain the knowledge, institutional support, and capacity necessary to exercise their rights and participate more equitably in social and economic systems.[11] For indigenous enterprises, legal empowerment may include awareness of intellectual property protection, digital commerce regulations, fair trade practices, consumer rights, and culturally respectful branding and promotion. Integrating these dimensions into extension initiatives can strengthen not only entrepreneurial capabilities but also community resilience, self-determination, and protection against exploitation.

---

7 Justine Lindemann et al., "Building Urban Community Resilience through University Extension: Community Engagement and the Politics of Knowledge," *Socio-Ecological Practice Research* 4, no. 4 (December 2022): 325–37, https://doi.org/10.1007/s42532-022-00126-6; André Lucas Novaes, "Participatory and Experiential Learning Methodologies in University Extension Programs: Integration Proposal Using an Application to Promote Climate Action Initiatives and Active Learning," *International Journal of Sustainability in Higher Education* 26, no. 6 (August 2025): 1267–83, https://doi.org/10.1108/IJSHE-08-2024-0525.

8 Mohammed Abdullatif Almulla, "Constructivism Learning Theory: A Paradigm for Students' Critical Thinking, Creativity, and Problem Solving to Affect Academic Performance in Higher Education," *Cogent Education* 10, no. 1 (December 2023): 2172929, https://doi.org/10.1080/2331186X.2023.2172929.

9 Jovelin Lapates and Ronel Bongga, "Bridging the Local Digital Divide of LGU Cabanglasan Towards an ICT-Enabled Community: Its Effectiveness," *Science International-Lahore* 34, no. 6 (2022): 493–97; Jovelin Lapates and Rozanne Tuesday Flores, "Digital Competence Level: An Evaluation of Area Coordinators and Emergency Response (ACER) Team ICT Proficiency," *International Journal of Multidisciplinary Research and Publications* 2, no. 1 (2019): 15–19; A. Mohammadi, M. Omidi Najafabadi, and A. Poursaeed, "A Comprehensive Sustainable Development Framework; Community Capitals and Village-Cooperative Initiative," *Brazilian Journal of Biology* 84 (2024): e269509, https://doi.org/10.1590/1519-6984.269509.

10 Golub, *Legal Empowerment: Practitioners' Perspectives*.

However, while numerous extension initiatives focus on skills training, there remains a lack of theoretically grounded and structured frameworks that integrate digital competency development with legal and community empowerment perspectives for indigenous communities. Existing studies on digital empowerment and community capacity-building often emphasize descriptive assessments of needs, challenges, and readiness levels.[12] Although these studies provide valuable baseline information, they frequently fall short of translating empirical findings into actionable educational and legal-community engagement models that can inform curriculum design, training pathways, policy development, and sustainable extension practices.[13]

This gap is particularly evident in the context of digital marketing education for indigenous enterprises, where the absence of structured and legally informed competency models limits the scalability, sustainability, and pedagogical coherence of extension interventions. Without a clearly defined framework, training programs risk being fragmented, tool-oriented, and disconnected from progressive learning pathways, legal awareness, and community-specific contexts. Furthermore, the absence of legal-literacy components may leave indigenous entrepreneurs vulnerable to exploitation, unfair market practices, and inadequate protection of cultural products and identity within digital spaces.

---

[11] Suhadi Suhadi et al., "Enhancing Nazir Capacity through Legal Literacy of Land Trust to Ensure Legal Certainty in Waqf Land Management," *Jurnal Pengabdian Hukum Indonesia* 7, no. 2 (2024), https://doi.org/10.15294/jphi.v7i2.6693.

[12] Bayu Dwi Anggono et al., "Interrogating the Legal Foundations of Digital Transformation: Balancing Economic Growth and Social Welfare in the Era of Disruption," *Volksgeist: Jurnal Ilmu Hukum Dan Konstitusi* 8, no. 1 (2025): 191–211, https://doi.org/https://doi.org/10.24090/volksgeist.v8i1.12211.

[13] Yang Liu et al., "Digital Transformation and Green Agricultural Skills: A Cross-Case Policy Analysis of Cost-Effective Competence Development," *European Journal of Education* 61, no. 2 (June 2026): e70537, https://doi.org/10.1111/ejed.70537; Teresa Ober et al., "Aligning Models of Competency-Based Education with Skills for the Future," *Chinese/English Journal of Educational Measurement and Evaluation* 6, no. 3 (September 2025), https://doi.org/10.59863/XIKM7863.

To address this gap, the present study reconceptualizes needs assessment not merely as a diagnostic exercise but as a foundation for legal-digital empowerment and framework development. Anchored on Market Development Theory,[14] the Capacity-Building Model,[15] the Sustainable Livelihoods Approach,[16] and Legal Empowerment Theory,[17] and informed by principles of experiential and community-based learning, this study utilizes empirical data from IPOs in the Philippines to systematically identify digital marketing and legal-literacy gaps and transform them into structured learning and empowerment domains.

Specifically, this study aims to: (1) analyze the current marketing practices and digital marketing competency levels of IPO members; (2) identify critical gaps across key domains such as content creation, digital platforms, branding and packaging, marketing strategy, and legal awareness; and (3) develop a LDEF-ICE that provides a structured and context-responsive model for designing capacity-building and legal-community engagement programs within higher education extension initiatives. By bridging empirical needs assessment with framework development, the study contributes to advancing educational practice and legal-community engagement by offering a scalable, rights-based, and theory-informed model that supports inclusive digital participation, legal empowerment, and sustainable indigenous enterprise development.

---

14 Benjamin Zugay and Rimi Zakaria, *Ansoff Matrix* (2455 Teller Road, Thousand Oaks California 91320 United States: SAGE Publications, Inc., 2023), https://doi.org/10.4135/9781071924136.

15 Petra T. Buergelt and Douglas Paton, "An Ecological Risk Management and Capacity Building Model," *Human Ecology* 42, no. 4 (August 2014): 591–603, https://doi.org/10.1007/s10745-014-9676-2.

16 R. Chambers and G. Conway, *Sustainable Rural Livelihoods: Practical Concepts for the 21st Century (Version 1)* (The Institute of Development Studies and Partner Organisations, 1992).

17 Golub, *Legal Empowerment: Practitioners' Perspectives*.

# Methods

## A. Research Design

This study employed a sequential mixed-methods design, integrating quantitative and qualitative approaches to provide a comprehensive understanding of the digital marketing competencies and capacity-building needs of IPO members. The quantitative component focused on measuring competency levels and identifying patterns in marketing practices, while the qualitative component provided deeper insights into contextual challenges and training needs.

The integration of these methods enabled triangulation of data, enhancing the validity and reliability of findings and supporting the development of a robust, evidence-based framework.[18] Furthermore, the design aligns with competency-based education approaches that emphasize the systematic structuring of skills into measurable domains and progressive levels.[19]

## B. Research Participants

The respondents of the study were members and community leaders from selected IPOs in the Bukidnon–Misamis Oriental landscape, Philippines. These organizations are engaged in various community-based enterprises, including bamboo crafts, woven textiles, food processing, and essential oil production.

A purposive sampling technique was employed to ensure that participants were directly involved in livelihood and marketing activities within their respective organizations. This method is appropriate in community-based research, as it allows for the selection of information-rich

---

[18] John W. Creswell and Vicki L. Plano Clark, "Revisiting Mixed Methods Research Designs Twenty Years Later," in *The Sage Handbook of Mixed Methods Research Design*, ed. Cheryl Poth (1 Oliver's Yard, 55 City Road London EC1Y 1SP: Sage Publications Ltd, 2023), 21–36, https://doi.org/10.4135/9781529614572.n6.

[19] Abid Haleem et al., "Understanding the Role of Digital Technologies in Education: A Review," *Sustainable Operations and Computers* 3 (2022): 275–85, https://doi.org/10.1016/j.susoc.2022.05.004.

participants who can provide relevant and meaningful insights into the study.[20]

## C. Research Instruments

The study utilized a structured survey questionnaire and a focus group discussion (FGD) guide as primary data collection instruments.

The survey questionnaire was designed to capture multiple dimensions of the study, including organizational profile, current marketing practices, marketing challenges, capacity-building needs, and digital marketing competencies. Competency levels were measured using a five-point proficiency scale: beginner, novice, intermediate, advanced, and expert, covering domains such as content creation, digital platforms, social media management, branding, packaging, and marketing strategy.

To complement the quantitative data, FGDs were conducted to gather in-depth insights into participants' experiences, perceptions, and contextual challenges. This combination of instruments enabled data triangulation and enhanced the validity of the findings.[21]

## D. Data Collection Procedure

Prior to data collection, ethical clearance was obtained from the Research Ethics Committee of Bukidnon State University. Coordination with partner organizations was conducted to facilitate access to respondents and ensure smooth implementation of the study.

Data were collected through both online and face-to-face survey administration, accommodating participants with varying levels of digital access. Respondents were informed about the purpose of the study, and informed consent was secured prior to participation.

Focus group discussions were conducted with selected IPO members to further explore their marketing practices, challenges, and training needs.

---

[20] Lawrence A. Palinkas et al., "Purposeful Sampling for Qualitative Data Collection and Analysis in Mixed Method Implementation Research," *Administration and Policy in Mental Health and Mental Health Services Research* 42, no. 5 (September 2015): 533–44, https://doi.org/10.1007/s10488-013-0528-y.

[21] Creswell and Plano Clark, "Revisiting Mixed Methods Research Designs Twenty Years Later."

All responses were recorded, transcribed, and systematically organized for analysis.

## E. Data Analysis

Quantitative data were analyzed using descriptive statistics, including frequencies and percentages, to determine the distribution of marketing practices, challenges, and competency levels among IPO members.

Qualitative data from FGDs were analyzed using thematic analysis, which involved coding responses, identifying recurring patterns, and organizing them into meaningful themes related to marketing challenges and capacity-building needs.

Beyond descriptive analysis, the study employed a competency mapping and framework synthesis approach, where identified gaps were systematically categorized into key domains—content creation, digital platforms, branding and packaging, and marketing strategy—and organized into progressive proficiency levels. This enabled the transformation of empirical findings into a structured competency framework aligned with educational objectives.[22]

## F. Framework Development Procedure

The development of LDEF-ICE followed a structured, iterative, and theory-informed process that integrated empirical findings, legal empowerment perspectives, and higher education extension principles. The procedure consisted of the following stages:

1. Identification of digital competency and legal-literacy gaps based on quantitative and qualitative findings, including challenges related to digital marketing practices, online participation, branding, consumer-related concerns, and awareness of legal protections relevant to indigenous enterprises;

[22] Simon F. J. Child and Stuart D. Shaw, "A Purpose-Led Approach towards the Development of Competency Frameworks," *Journal of Further and Higher Education* 44, no. 8 (September 2020): 1143–56, https://doi.org/10.1080/0309877X.2019.1669773.

2. Clustering of findings into interconnected thematic domains, including digital enterprise competencies, legal literacy and rights awareness, institutional support systems, and sustainable market participation;
3. Classification of progressive proficiency levels (beginner, novice, intermediate, and advanced) to establish scaffolded learning pathways and competency progression consistent with competency-based education approaches;
4. Integration of legal and community empowerment components such as intellectual property awareness, consumer protection, e-commerce participation, culturally respectful branding, and protection of indigenous identity and products within digital environments;
5. Synthesis of learning pathways, intervention strategies, and support mechanisms, including mentorship programs, legal orientation activities, experiential learning opportunities, and community-based extension initiatives;
6. Alignment of the framework with theoretical foundations and higher education extension practices, particularly Legal Empowerment Theory, Market Development Theory, the Capacity-Building Model, the Sustainable Livelihoods Approach, experiential learning, and community-based education principles.

This process ensured that the framework is empirically grounded, contextually responsive, legally informed, and pedagogically sound. Furthermore, the procedure supports the development of extension and community engagement programs that integrate digital inclusion, legal literacy, and rights-based participation, making the framework suitable for application in higher education extension and legal-community engagement initiatives involving indigenous communities.

# Results and Discussion

## A. Current Marketing Practices and Strategies

The survey results provide insight into how IPOs in the Philippines currently market their products as shown in Table 1.

**TABLE 1**

Current Marketing Strategies of IP Organizations

| Category | Frequency |
|---|---|
| **Sales Channels** | |
| Local Markets | 7 |
| Trade Fairs & Exhibitions | 12 |
| Partner Retail Stores | 6 |
| Online Platforms | 5 |
| Cooperatives | 18 |
| Others | 1 |
| **Branding** | |
| Complete Branding & Packaging System | 7 |
| Inconsistent or Incomplete Branding | 6 |
| No Branding | 2 |
| **Marketing Materials** | |
| Poster & Brochures | 1 |
| Product Catalogs | 1 |
| None of the Above | 18 |
| **Marketing Management** | |
| Individual | 3 |
| Assigned Team | 18 |
| External Partners | 7 |
| No One Assigned | 1 |
| **Pricing** | |
| Based on Production Cost & Markup | 18 |
| Based on Customers' Willingness to Pay | 0 |
| Based Competitor's Prices | 0 |
| No Clear Pricing Strategy | 1 |

The findings indicate that IPOs predominantly rely on traditional marketing channels, particularly cooperatives, trade fairs, and local markets, with minimal engagement in digital platforms. The limited adoption of online marketing reflects a significant gap in digital readiness and highlights the need for foundational competencies in digital marketing.

Branding and marketing materials are also underdeveloped, with a substantial number of IPOs lacking consistent branding systems and formal promotional materials. Although marketing responsibilities are often assigned within organizations, strategic marketing practices such as customer-based pricing and competitive positioning remain largely absent.

These findings suggest that IPOs operate within a traditional marketing paradigm, where digital tools and strategic marketing concepts are underutilized. This condition underscores the need to develop competencies not only in digital tools but also in strategic marketing thinking, which becomes a critical component of the proposed framework.

## B. Capacity-Building Needs and Readiness

The survey results on capacity-building needs reveal a strong desire among IPOs to enhance their marketing knowledge and skills as presented in Table 2.

**TABLE 2**

Capacity-Building Needs of IP Organizations

| Feedback | Mentions |
|---|---|
| **Marketing Skills/Training Needs** | |
| Branding & packaging design | 18 |
| Pricing & financial management | 13 |
| Online marketing | 13 |
| Customer engagement & storytelling | 10 |
| Legal aspects | 15 |
| Other | 0 |
| **Willingness to Participate in Training Workshops** | |
| Yes | 18 |
| No | 0 |
| **Access to Digital Tools for Marketing** | |
| Yes, we use them regularly | 2 |
| Sometimes, but access is limited | 16 |
| No, we do not have access | 0 |
| **Interest in Mentorship** | |
| Yes | 18 |
| No | 0 |

The findings demonstrate a strong willingness among IPO members to engage in training and mentorship programs, with all respondents expressing interest in capacity-building initiatives. Key areas identified include branding and packaging, online marketing, pricing strategies, legal aspects and customer engagement.

Despite this readiness, access to digital tools remains limited, with only a small proportion of respondents regularly using devices and internet connectivity for marketing purposes. This highlights a critical gap between learning readiness and resource availability, suggesting that training programs must be designed to be accessible, scalable, and adaptable to low-resource environments.

From a framework perspective, these results emphasize the importance of integrating progressive learning pathways and tool accessibility considerations into the competency model.

## C. Participants Feedback

Table 3 shows the feedback provided by the IP members during needs assessment. The most frequently suggested intervention by IPO members is to enable online selling and internet-based promotion, mentioned by four respondents. This reflects a strong awareness among IPO members of the growing importance of digital platforms in expanding market reach, particularly for communities with limited access to physical markets. The demand for improved branding, packaging, and product design was also evident, mentioned by three respondents, indicating recognition that product presentation and visual identity play a key role in attracting customers.

**TABLE 3**
Feedback from IP Members

| Feedback | Mentions |
|---|---|
| **Ways to Improve Marketability of Products** | |
| Enable online selling & internet-based promotion | 4 |
| Trainings & mentoring | 2 |
| Improve branding, packaging, and product design | 3 |
| Social media marketing & posting | 2 |
| Product quality improvement | 2 |
| Modern communication training | 1 |
| **Additional Comments and Suggestions** | |
| Project is a good idea to improve business knowledge | 1 |
| Difficulty sustaining business due to lack of capital | 1 |
| Lack of production equipment | 1 |
| Lack of legal awareness | 1 |
| Desire for more marketing training | 1 |
| Hope for additional funding | 1 |

Additionally, there is significant interest in trainings and mentoring with 2 mentions, as well as social media marketing and posting with 2 mentions, both of which point to the IPO members' desire for capacity-building and hands-on support to navigate modern marketing tools. Other suggestions include improving overall product quality with 2 mentions and modern communication training with 1 mention, reflecting the need to ensure products meet market expectations and that members are equipped with communication skills for customer engagement.

## D. Digital Marketing Skills on Publication Materials

The results of the digital marketing needs assessment for IPOs in the Philippines reveal critical gaps in design and visual content creation competencies, particularly in producing posters, infographics, and promotional materials essential for improving market visibility as shown in Table 4.

**TABLE 4**

Percentage Distribution of Digital Marketing Skills on Posters and Infographics

| Competency | % Beginner | % Novice | % Intermediate | % Advanced | % Expert |
|---|---|---|---|---|---|
| Design visually appealing posters and infographics for digital campaigns. | 66.67 | 22.22 | 11.11 | 0 | 0 |
| Use design tools. | 77.78 | 11.11 | 11.11 | 0 | 0 |
| Apply branding guidelines. | 77.78 | 16.67 | 5.56 | 0 | 0 |
| Optimize images for different social media platforms. | 88.89 | 11.11 | 0 | 0 | 0 |
| Create engaging, informative, and shareable content. | 77.78 | 16.67 | 5.56 | 0 | 0 |
| Understand basic design principles. | 88.89 | 11.11 | 0 | 0 | 0 |
| Use beginner-friendly tools to create simple posters. | 88.89 | 11.11 | 0 | 0 | 0 |
| Create infographics using templates with minimal customization. | 88.89 | 11.11 | 0 | 0 | 0 |
| Learn video formats and dimensions for different platforms. | 88.89 | 11.11 | 0 | 0 | 0 |
| Use mobile-friendly tools to edit short videos. | 88.89 | 0 | 11.11 | 0 | 0 |
| Add text, transitions, and background music to videos. | 83.33 | 5.56 | 11.11 | 0 | 0 |
| Use Adobe Premiere Pro, After Effects, or DaVinci Resolve for basic video editing. | 88.89 | 11.11 | 0 | 0 | 0 |
| Create branded promotional videos. | 77.78 | 22.22 | 0 | 0 | 0 |
| Add motion graphics and simple animations to product videos. | 77.78 | 22.22 | 0 | 0 | 0 |
| Customize poster and infographic templates with unique brand elements. | 88.89 | 11.11 | 0 | 0 | 0 |
| Apply brand consistency across marketing materials. | 94.44 | 5.56 | 0 | 0 | 0 |
| Design promotional banners for digital ads. | 77.78 | 22.22 | 0 | 0 | 0 |
| Design high-converting ad creatives for multiple platforms. | 83.33 | 11.11 | 5.56 | 0 | 0 |
| Implement Alpha and Beta testing to optimize ad designs. | 83.33 | 16.67 | 0 | 0 | 0 |
| Create interactive infographics for websites & social media. | 77.78 | 22.22 | 0 | 0 | 0 |
| Produce high-quality promotional videos with cinematic effects. | 83.33 | 11.11 | 5.56 | 0 | 0 |
| Create product explainer animations using motion graphics. | 88.89 | 5.56 | 5.56 | 0 | 0 |
| Use Artificial Intelligence (AI)-driven tools for automated video editing & enhancement. | 83.33 | 11.11 | 5.56 | 0 | 0 |
| Create infographics using templates with minimal customization. | 88.89 | 11.11 | 0 | 0 | 0 |

The data reveals a widespread lack of proficiency among IPO members in designing posters and infographics, with beginner-level self-ratings across all 24 assessed indicators ranging from 66.67% to 88.89%. Basic skills such as creating visually appealing posters, using tools like Canva or Photoshop, and applying branding guidelines showed beginner rates of up to 77.78%, while more technical tasks like content optimization for social media, ensuring brand consistency, customizing infographics, and applying branding to digital ads reflected even higher beginner rates, with virtually no respondents identifying as advanced users.

The findings underscore the importance of providing structured, skill-appropriate training that starts with the basics of digital design, progressing to branding application, content optimization, and eventually, more advanced techniques in promotional material creation.

Beyond these technical competency gaps, the findings also reveal significant legal implications for IPOs. As members begin producing posters, promotional materials, logos, and digital content without adequate legal literacy, they may unknowingly expose indigenous cultural symbols, traditional designs, community knowledge, and product identities to unauthorized reproduction, commercial exploitation, or misappropriation in online environments. Limited awareness of copyright, trademark protection, and intellectual property rights may prevent IPOs from securing legal ownership of their creative outputs or asserting their rights when cultural products are reproduced or commercialized without prior consent, proper attribution, or equitable benefit-sharing. These findings suggest that training on digital content creation should be complemented by legal education that enables indigenous communities to protect both their economic interests and their cultural heritage.

**TABLE 5**

Percentage Distribution of Digital Marketing Skills on Video and Short Product Animation

| **Competency** | **% Beginner** | **% Novice** | **% Intermediate** | **% Advanced** | **% Expert** |
|---|---|---|---|---|---|
| Edit and produce engaging promotional videos. | 77.78 | 11.11 | 11.11 | 0 | 0 |
| Create short product animations using | 77.78 | 22.22 | 0 | 0 | 0 |
| After Effects, Canva, or AI tools. | | | | | |
| Optimize video formats for different platforms. | 66.67 | 22.22 | 11.11 | 0 | 0 |
| Implement branding elements into video content. | 88.89 | 11.11 | 0 | 0 | 0 |

The assessment results for video production and short product animation reveal a consistent trend of low proficiency among IPOs in utilizing this crucial digital marketing tool as illustrated in Table 5. Across all four indicators under this category, many respondents self-rated their skills at the beginner level, underscoring significant capacity gaps that hinder their ability to create engaging, market-ready video content.

The data highlights a significant lack of video production and animation skills among IPO members, with 77.78% identifying as beginners in editing promotional videos and creating short product animations using tools like After Effects, Canva, or AI platforms, while no respondents reported advanced proficiency. Similarly, 66.67% lack the technical skills to optimize video formats for different platforms, with only 11.11% reaching intermediate competency. Most notably, 88.89% reported beginner-level ability to integrate branding elements into video content, underscoring a critical gap in using video not only for promotion but also for building brand identity, market recognition, and consumer trust—an essential requirement in today's digital marketing landscape.

Collectively, these results demonstrate that IPO members currently lack the technical expertise and confidence to produce high-quality, branded video content that effectively showcases their products and tells their unique stories. The absence of advanced skills and the overwhelming concentration of respondents at the beginner level point to the need for structured, hands-on training in video editing, animation, and branding integration.

## E. Digital Marketing Skills on Digital Marketing Mediums

The assessment of IPO members' proficiency in digital marketing mediums reveals substantial gaps in both technical skills and strategic understanding, posing significant barriers to effectively promoting indigenous products in the digital marketplace. As shown in Table 6, the results from the 14 assessed indicators consistently demonstrate that most respondents self-identified as beginners, while advanced and expert-level skills remain virtually absent across all competencies.

**TABLE 6**

Percentage Distribution of Digital Marketing Skills on Digital Marketing Mediums

| Competency | % Beginner | % Novice | % Intermediate | % Advanced | % Expert |
|---|---|---|---|---|---|
| Understanding different digital marketing channels. | 77.78 | 22.22 | 0 | 0 | 0 |
| Choosing the right medium for a specific marketing goal. | 72.22 | 22.22 | 5.56 | 0 | 0 |
| Measuring performance and ROI of digital campaigns. | 83.33 | 16.67 | 0 | 0 | 0 |
| Using tools like Google Ads, Facebook Ads, and email automation. | 83.33 | 11.11 | 5.56 | 0 | 0 |
| Creating a multi-channel marketing strategy. | 88.89 | 5.56 | 5.56 | 0 | 0 |
| Understand the difference between organic and paid digital marketing. | 88.89 | 11.11 | 0 | 0 | 0 |
| Learn the basics of content marketing and digital advertising. | 88.89 | 11.11 | 0 | 0 | 0 |
| Set up Google Ads & Facebook Ads Manager for basic ad campaigns. | 88.89 | 11.11 | 0 | 0 | 0 |
| Use email marketing tools to create campaigns. | 88.89 | 11.11 | 0 | 0 | 0 |
| Learn Search Engine Optimization (SEO) basics. | 83.33 | 11.11 | 5.56 | 0 | 0 |
| Plan and execute multi-channel campaigns. | 88.89 | 5.56 | 5.56 | 0 | 0 |
| Optimize landing pages for higher conversion rates. | 88.89 | 11.11 | 0 | 0 | 0 |
| Automate marketing workflows using HubSpot, Zapier, or ActiveCampaign | 88.89 | 11.11 | 0 | 0 | 0 |
| Identify different digital marketing channels. | 88.89 | 11.11 | 0 | 0 | 0 |

The data reveals a widespread lack of digital marketing knowledge and strategic awareness among IPO members, with 77.78% identifying as beginners in understanding various digital marketing channels like SEO (Search Engine Optimization), PPC (Pay-Per-Click), social media, and email marketing, and only 22.22% reaching the novice level. Similar gaps exist in selecting appropriate marketing platforms, measuring ROI, using tools like Google Ads and Facebook Ads, and setting up automation platforms, where beginner rates consistently exceed 83%. Technical and strategic competencies such as developing multi-channel marketing strategies, distinguishing between organic and paid efforts, content marketing fundamentals, SEO basics, landing page optimization, and automating marketing workflows also show beginner-level dominance, with no respondents reporting advanced or expert proficiency. These results highlight the urgent need for foundational capacity-building programs to equip IPO members with essential digital marketing skills, improve product visibility, and enhance the long-term competitiveness and sustainability of their livelihood initiatives.

Moreover, participation in digital marketplaces introduces legal responsibilities and risks that extend beyond marketing competence. Limited understanding of e-commerce regulations, consumer protection laws, digital platform policies, data privacy requirements, and intellectual property protection may leave IPO members vulnerable to unfair commercial practices and misuse of their indigenous products and traditional knowledge. Strengthening legal literacy alongside digital marketing competencies is therefore essential to ensure that indigenous enterprises can participate confidently in digital economies while safeguarding their cultural identity and exercising their legal rights.

## F. Digital Marketing Skills on Social Media Platforms

Table 7 presents the percentage distribution of IPO members' self-assessed proficiency in managing social media platforms, highlighting considerable gaps in both operational and strategic competencies required for effective online engagement.

**TABLE 7**

Percentage Distribution of Digital Marketing Skills on Social Media Platforms

| Competency | % Beginner | % Novice | % Intermediate | % Advanced | % Expert |
|---|---|---|---|---|---|
| Managing business accounts on platforms like Facebook, Instagram, LinkedIn, TikTok, and YouTube. | 83.33 | 11.11 | 5.56 | 0 | 0 |
| Understanding algorithms and engagement strategies for each platform. | 83.33 | 11.11 | 5.56 | 0 | 0 |
| Creating platform-specific content. | 61.11 | 27.78 | 11.11 | 0 | 0 |
| Using social media analytics tools. | 77.78 | 16.67 | 5.56 | 0 | 0 |
| Running paid advertising campaigns on social media. | 83.33 | 11.11 | 5.56 | 0 | 0 |
| Create and manage basic business accounts on Facebook, Instagram, LinkedIn, and TikTok. | 77.78 | 16.67 | 5.56 | 0 | 0 |
| Post static images and text-based content. | 61.11 | 27.78 | 11.11 | 0 | 0 |
| Understand the fundamentals of hashtags and basic engagement strategies. | 88.89 | 0 | 11.11 | 0 | 0 |
| Develop a 30-day social media content calendar. | 88.89 | 5.56 | 5.56 | 0 | 0 |
| Use social media insights & analytics to track performance. | 88.89 | 0 | 11.11 | 0 | 0 |
| Engage with audiences via comments and community posts. | 66.67 | 27.78 | 5.56 | 0 | 0 |
| Implement retargeting strategies for paid social media ads. | 88.89 | 5.56 | 5.56 | 0 | 0 |
| Run influencer collaborations & social media partnerships. | 44.44 | 55.56 | 0 | 0 | 0 |
| Use advanced analytics & social listening tools. | 88.89 | 5.56 | 5.56 | 0 | 0 |

The data reveals widespread skill gaps in social media management among IPO members, with 83.33% identifying as beginners in managing business accounts on platforms like Facebook, Instagram, LinkedIn, TikTok, and YouTube, and the same percentage reporting beginner-level understanding of platform-specific algorithms and engagement strategies. Content creation for these platforms shows slightly better results, yet 61.11% remain at the beginner level, with only 11.11% reaching intermediate proficiency. Similarly, 77.78% of respondents lack proficiency in using analytics tools, and 83.33% are beginners in running paid advertising campaigns. Across all 14 social media indicators, no respondents possess advanced or expert-level skills, reflecting significant limitations in high-level social media management, including content planning, audience engagement, and retargeting strategies necessary for effective digital marketing.

These findings reveal that IPO members have a very limited capacity to use social media as a powerful tool for business growth, customer engagement, and product promotion.

## G. Synthesis: From Needs Assessment to Legal-Digital Empowerment Framework

The integration of findings across all competency domains reveals that the challenges faced by IPOs extend beyond deficiencies in digital marketing skills and represent broader legal and institutional vulnerabilities affecting their participation in digital economies. While the predominance of beginner-level competencies in digital communication, branding, online promotion, and content creation indicates the need for structured learning pathways, these limitations also reduce the capacity of IPOs to protect indigenous cultural expressions, traditional knowledge, community brands, and creative works within digital environments. Consequently, limited digital competence should be understood not only as a technical constraint but also as a legal vulnerability that may expose indigenous communities to unauthorized reproduction, cultural appropriation, intellectual property infringement, and commercial exploitation without appropriate legal protection or equitable benefit-sharing.

More specifically, the findings indicate that inadequate legal literacy increases the vulnerability of indigenous cultural products, traditional knowledge, creative works, and community brands to unauthorized use in digital environments. As IPO members expand their online presence through digital marketing and social media, they may unintentionally disclose culturally significant knowledge or creative expressions without understanding the legal mechanisms available to protect these assets. Such limitations increase the risk of copyright infringement, cultural appropriation, unauthorized commercialization, and the loss of potential economic benefits that should rightfully accrue to indigenous communities. Consequently, legal literacy should be regarded as a fundamental component of digital empowerment, enabling IPO members not only to market their products effectively but also to protect their intellectual and cultural heritage while participating equitably in digital economies.

The synthesis of the findings therefore reconceptualizes the needs assessment as a process of identifying legal and digital vulnerabilities that require integrated empowerment strategies. Rather than viewing digital marketing competency as an isolated technical objective, the study recognizes that sustainable participation in digital markets requires simultaneous development of digital competencies, legal literacy, and institutional support. Without adequate understanding of intellectual property rights, copyright protection, trademarks, consumer protection regulations, and digital platform governance, indigenous enterprises may unintentionally disclose or commercialize valuable cultural assets without sufficient legal safeguards. Accordingly, digital competency development should be pursued alongside legal empowerment to strengthen indigenous communities' ability to protect their cultural heritage, assert their rights, and participate equitably in digital economies. This integrated approach also depends on the availability of secure and reliable digital infrastructures that enable communities to participate safely in digital environments. As emphasized by Cagande et al.,[23] sustainable digital transformation requires technological infrastructures that integrate security, scalability, and adaptability to support long-term digital engagement.

The findings also highlight the importance of scaffolded, context-sensitive, and legally informed learning pathways. Since many IPO members remain at the beginner and novice stages of digital participation, extension programs should progressively develop both digital competencies and legal awareness. Such an approach enables participants not only to create and market digital content effectively but also to understand the legal implications of publishing indigenous cultural products and traditional knowledge in online environments. Integrating legal literacy into competency development equips IPO members with the knowledge necessary to safeguard intellectual property, prevent cultural misappropriation, comply with relevant legal frameworks, and engage responsibly in digital commerce.

The findings further suggest that legal literacy should be integrated into digital competency development rather than delivered as a stand-alone intervention. Such an integrated approach enables participants to move beyond acquiring technical skills toward applying legal knowledge in ways that support the protection of indigenous cultural assets and meaningful participation in digital economies. This perspective is consistent with previous research emphasizing that awareness should be reinforced through active stakeholder engagement and opportunities for practical application to achieve sustained participation and alignment.[24] Moreover, the synthesis underscores the strategic role of HEIs in facilitating community-based legal and digital empowerment initiatives. Through extension and engagement programs, HEIs can serve as institutional partners that provide training, legal orientation, mentorship, research support, and collaborative learning opportunities for indigenous communities. This expands the role of higher education beyond traditional knowledge transfer toward participatory and socially responsive community engagement that addresses both educational and legal dimensions of development.[25]

---

[23] Gil Nicholas Cagande et al., "Network Design for Bukidnon State University College of Technologies Annex B Campus," *International Journal of Engineering Trends and Technology* 72, no. 11 (November 2024): 207–44, https://doi.org/10.14445/22315381/IJETT-V72I11P122.

[24] Sales Jr. Aribe, Marilou Espina, and Joan Marie Panes, "Stakeholder Perceptions on VMGO, Institutional Outcomes and Core Values in Creative Computing Education in the Philippines: Awareness, Acceptance, and Congruency," *International Journal of Education and Practice* 14, no. 1 (January 2026): 337–53, https://doi.org/10.18488/61.v14i1.4720.

Overall, the synthesis transforms the needs assessment into a framework-building process in which empirical findings inform the development of competency domains, legal-literacy components, institutional support mechanisms, and sustainable rights-based participation strategies. The integration of these elements provides the conceptual and empirical foundation for the proposed Legal-Digital Empowerment Framework for Indigenous Community Enterprises (LDEF-ICE), presented in the succeeding section.

## H. Proposed Legal-Digital Empowerment Framework for Indigenous Community Enterprises

Building on the empirical findings of the needs assessment and competency mapping process, this study proposes LDEF-ICE as a structured and community-based model for strengthening the digital participation, legal literacy, and sustainable market engagement of IPOs. Rather than limiting capacity-building to technical digital marketing competencies, the framework integrates legal empowerment, community education, and rights-based participation as essential components of indigenous enterprise development within higher education extension programs.

The framework, as shown in Figure 1, is anchored on the integration of Legal Empowerment Theory, the Sustainable Livelihoods Approach, the Capacity-Building Model, and community-based educational perspectives. Legal empowerment emphasizes that marginalized communities can improve their socio-economic conditions when they are equipped with legal knowledge, institutional access, and the capacity to exercise their rights in economic and social systems.[26] Within the context of indigenous enterprises, this includes awareness of intellectual property rights, consumer protection regulations, digital platform policies, fair trade principles, and legal mechanisms that protect indigenous cultural identity and products from exploitation.[27] The Sustainable Livelihoods Approach complements this perspective by emphasizing that sustainable development depends not only on economic resources but also on social, institutional, and human capacities that enable communities to participate equitably in markets and governance structures.[28] Meanwhile, the Capacity-Building Model highlights the importance of progressively developing competencies, organizational systems, and institutional support necessary for long-term empowerment and resilience.[29]

[25] Corina Joseph and Roshima Said, "Community-Based Education: A Participatory Approach to Achieve the Sustainable Development Goal," in *Quality Education*, ed. Walter Leal Filho et al. (Cham: Springer International Publishing, 2020), 101–11, https://doi.org/10.1007/978-3-319-95870-5_67; Lindemann et al., "Building Urban Community Resilience through Univer."

[26] Golub, *Legal Empowerment: Practitioners' Perspectives*.

These theoretical foundations are further strengthened through experiential learning and community-based education approaches, which position HEIs as facilitators of collaborative and practice-oriented learning.[30] Through extension and community engagement programs, HEIs serve as partners in promoting legal literacy, digital inclusion, and community empowerment among IPOs.[31]

The development of the framework followed an iterative synthesis process in which quantitative and qualitative findings were systematically transformed into interconnected learning and empowerment domains. Competency gaps, legal concerns, and structural barriers identified during the needs assessment were clustered into thematic categories involving legal literacy, digital enterprise competencies, institutional support systems, and sustainable market participation. These categories were subsequently organized into progressive proficiency levels and integrated with legal empowerment and community-based education perspectives to establish structured learning pathways and intervention mechanisms. Through this process, the framework operationalizes empirical findings into a context-responsive model for higher education extension and legal-community engagement initiatives.

---

27 Anindya Bhukta, *Legal Protection for Traditional Knowledge: Towards a New Law for Indigenous Intellectual Property* (United Kingdom: Emerald Publishing Limited, 2020).

28 Chambers and Conway, *Sustainable Rural Livelihoods: Practical Concepts for the 21st Century (Version 1)*.

29 Buergelt and Paton, "An Ecological Risk Management and Capacity Building Model."

30 David Allen Kolb, *Experiential Learning: Experience as the Source of Learning and Development*, 1984; Joseph and Said, "Community-Based Education."

31 Lindemann et al., "Building Urban Community Resilience through Univer"; Novaes, "Participatory and Experiential Learning Methodolog."

The structure of the LDEF-ICE consists of four interconnected components: legal literacy and rights awareness, digital enterprise competencies, community-based support systems, and sustainable rights-based market participation.

The first component, legal literacy and rights awareness, focuses on strengthening IPO members' understanding of legal concepts directly related to community enterprises and digital participation. Findings from the study revealed that legal aspects emerged as one of the most frequently identified training needs among respondents, indicating a growing recognition that successful participation in digital markets also requires legal awareness and protection mechanisms. Under this component, IPO members are introduced to concepts such as intellectual property rights,[32] indigenous cultural protection, consumer rights, fair pricing practices, product labelling compliance, digital platform regulations, data privacy, and basic e-commerce policies.[33] This component recognizes that digital inclusion is not solely a technical issue but also a matter of legal access, protection, and social justice.[34]

The second component, digital enterprise competencies, encompasses the development of practical skills in digital communication, branding, content creation, online marketing, and social media management. Derived from the competency gaps identified in the needs assessment, this component organizes learning into progressive proficiency levels: beginner, novice, intermediate, and advanced. At the beginner level, learners develop foundational awareness of digital tools and online communication. The novice level focuses on guided application of digital marketing practices such as creating social media content and using beginner-friendly design platforms. Intermediate competencies involve independent management of online campaigns, audience engagement, and strategic branding, while advanced competencies emphasize adaptive and data-driven digital marketing practices. This scaffolded structure supports gradual learning progression consistent with competency-based education models.[35]

---

[32] Richard A. Spinello, "Intellectual Property Rights," in *Encyclopedia of Libraries, Librarianship, and Information Science* (Elsevier, 2025), 306–13, https://doi.org/10.1016/B978-0-323-95689-5.00134-6; Carolina Castaldi et al., "Are Intellectual Property Rights Working for Society?," *Research Policy*, 2024, https://doi.org/10.1016/j.respol.2023.104936.

[33] Jack Trọng, "Relationship between Data Privacy Regulations and Consumer Trust in E-Commerce Platforms in Vietnam," *American Journal of Data, Information and Knowledge Management* 5, no. 1 (2024), https://doi.org/10.47672/ajdikm.2347; Shiqiang Wang et al., "Security Risks and Regulation of Cross-Border e-Commerce in Digital Economy," *Journal of Electronic Business & Digital Economics* 4, no. 2 (2025), https://doi.org/10.1108/jebde-12-2024-0053.

[34] United Nations, "United Nations Declaration on the Rights of Indigenous Peoples. Resolution 61/295."

The third component focuses on community-based legal and institutional support systems. The findings indicate that IPO members face structural barriers such as limited infrastructure, inadequate access to digital tools, lack of mentorship, and insufficient institutional assistance. To address these issues, the framework positions HEIs, local government units, non-government organizations, and community stakeholders as collaborative support networks that facilitate sustained capacity-building. Within this component, extension programs may include legal education seminars, mentorship programs, digital marketing workshops, policy orientation activities, and technical assistance for online business participation. In this context, HEIs function not only as training providers but also as community partners that promote legal empowerment and inclusive development through participatory engagement.[36]

The fourth component, sustainable rights-based market participation, represents the long-term outcome of the framework. Through the integration of legal literacy, digital competencies, and institutional support, IPO members are expected to participate more effectively and responsibly in digital markets while preserving their cultural identity and protecting community interests. This component emphasizes ethical digital promotion, culturally respectful branding, lawful participation in e-commerce platforms, and equitable access to economic opportunities. It further highlights the importance of empowering indigenous communities to exercise agency over their products, narratives, and market relationships within increasingly digital economies.[37]

[35] Alejandro Curado Fuentes and Mercedes Rico García, "Digital Competence Analysis to Inform Current Pedagogy and Curricula," *Research in Education and Learning Innovation Archives*, no. 33 (July 2024), https://doi.org/10.7203/realia.33.29257; Haleem et al., "Understanding the Role of Digital Technologies In ."

[36] Lindemann et al., "Building Urban Community Resilience through Univer"; Mohammadi, Omidi Najafabadi, and Poursaeed, "A Comprehensive Sustainable Development Framework; Community Capitals and Village-Cooperative Initiative."

A defining feature of the LDEF-ICE is its integration within higher education extension and community engagement initiatives. By embedding legal and digital empowerment into extension programs, HEIs can transform community engagement into a holistic learning ecosystem that combines education, legal awareness, technological inclusion, and livelihood development. This aligns with contemporary perspectives on university extension, where institutions serve as active contributors to social justice, community resilience, and sustainable development [38].

The framework also carries important implications for policy and educational practice. First, it provides a structured basis for integrating legal literacy and digital entrepreneurship into extension curricula and community training programs. Second, it highlights the role of HEIs in supporting indigenous rights protection through community-based legal education and digital inclusion initiatives. Third, it demonstrates how interdisciplinary approaches combining law, education, and digital enterprise development can address broader social issues such as economic marginalization, unequal market participation, and limited access to digital opportunities among indigenous populations.

Overall, the LDEF-ICE advances the study beyond a conventional digital marketing framework by positioning competency development within a broader legal and community empowerment perspective. By integrating legal awareness, digital skills, institutional collaboration, and rights-based participation, the framework provides a theory-informed and actionable model for strengthening indigenous enterprises through higher education extension and community engagement programs.

---

37 Eric M. Olson et al., "Business Strategy and the Management of Digital Marketing," *Business Horizons* 64, no. 2 (March 2021): 285–93, https://doi.org/10.1016/j.bushor.2020.12.004.

38 Joseph and Said, "Community-Based Education"; Novaes, "Participatory and Experiential Learning Methodolog."

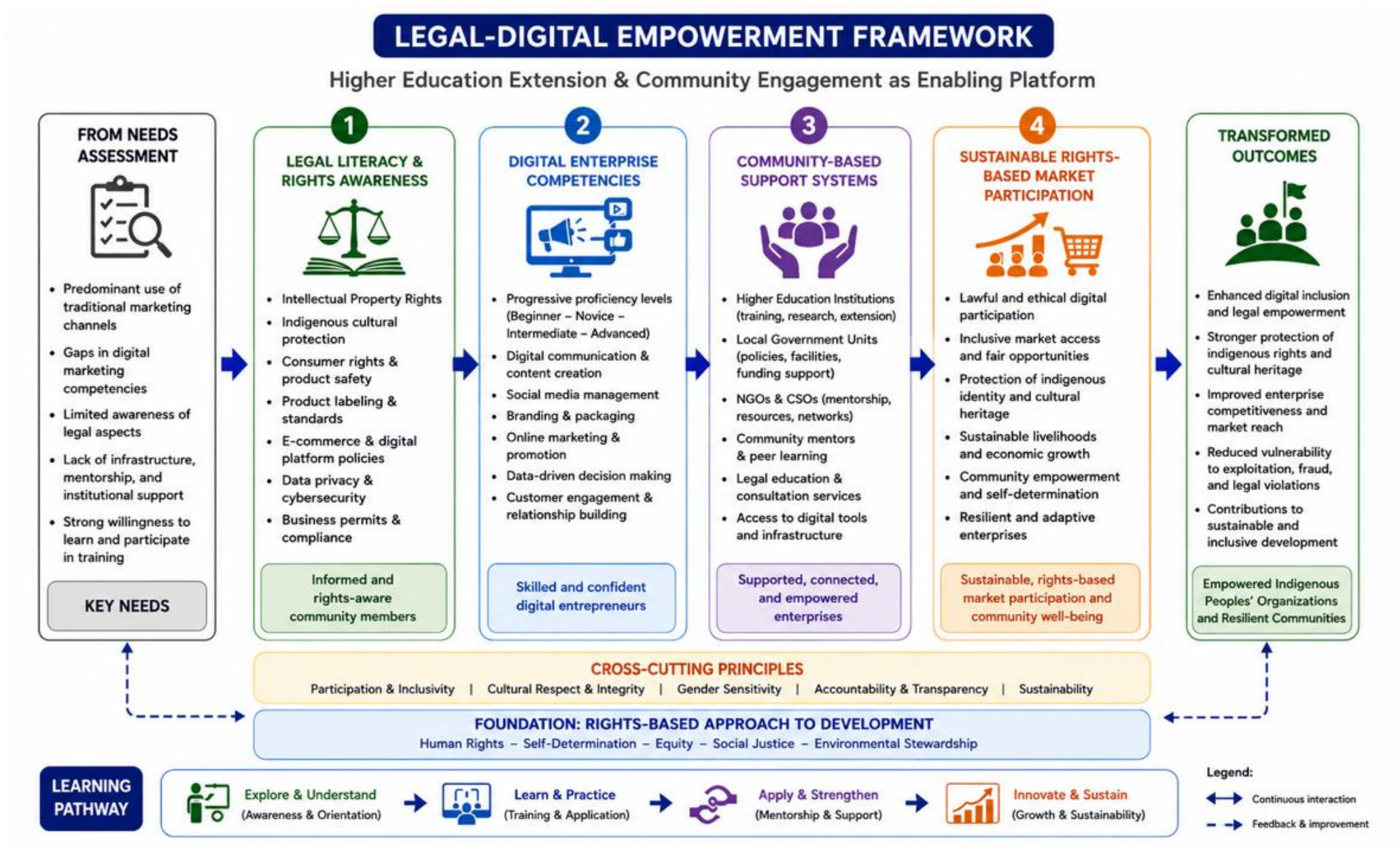

Figure 1. Proposed Legal-Digital Empowerment Framework for Indigenous Community Enterprises (LDEF-ICE)

## Conclusion

IPOs continue to play an important role in sustaining local livelihoods, preserving cultural identity, and supporting environmentally responsible community enterprises. However, the findings of the study reveal that many IPOs remain situated at the margins of digital economies due to limited digital competencies, inadequate access to technological resources, and insufficient legal awareness related to online enterprise participation. While traditional marketing approaches such as cooperatives, trade fairs, and local markets remain dominant among respondents, the adoption of digital marketing tools and strategic online engagement practices remains minimal. Branding systems, digital promotion strategies, and formal marketing structures also remain underdeveloped across many organizations.

The study further demonstrates that the challenges experienced by IPOs extend beyond technical skill limitations alone. Significant gaps were identified in content creation, social media management, branding and packaging, digital marketing strategy, and online communication. At the same time, respondents expressed concerns associated with legal and rights-based dimensions of enterprise participation, including awareness of intellectual property protection, digital platform participation, consumer-related practices, and protection of indigenous products and cultural identity within digital environments. These findings suggest that digital exclusion among indigenous enterprises is not solely a technological concern but also a legal and social justice issue connected to unequal access, vulnerability to exploitation, and limited participation in emerging digital economies.

The strong willingness of IPO members to engage in training, mentorship, and community-based learning initiatives highlights significant potential for legal and digital empowerment when supported by structured and context-sensitive interventions. The findings also emphasize the importance of institutional collaboration and sustained community support systems involving HEIs, local government units, and other stakeholders. Through extension and community engagement programs, HEIs are positioned to serve as facilitators of legal literacy, digital inclusion, entrepreneurial development, and participatory learning within indigenous communities.

Building on these findings, the study developed LDEF-ICE, which integrates legal literacy and rights awareness, digital enterprise competencies, community-based support systems, and sustainable rights-based market participation into a unified framework for indigenous empowerment. Anchored on Legal Empowerment Theory, Market Development Theory, the Capacity-Building Model, and the Sustainable Livelihoods Approach, the framework provides a structured and theory-informed approach for designing higher education extension and legal-community engagement initiatives. By integrating experiential learning, legal awareness, and digital competency development, the framework moves beyond conventional skills-training models toward a more holistic and community-centered approach to empowerment.

Overall, the study demonstrates that needs assessment can function not only as a mechanism for identifying competency gaps but also as a foundation for developing actionable and rights-based empowerment frameworks for marginalized communities. The proposed framework contributes to the advancement of higher education extension and legal-community engagement by offering a scalable and context-responsive model that supports digital inclusion, legal empowerment, sustainable enterprise participation, and community resilience among indigenous populations.

***

## Declaration of Conflicting Interests

The authors state that there is no conflict of interest in the publication of this article.

## Funding Information

None.

## Acknowledgment

The authors would like to express their sincere gratitude to Non-Timber Forest Products Exchange Programme (NTFP-EP) Philippines and the various IPOs in Bukidnon and Misamis Oriental for their invaluable participation, cooperation, and support throughout the conduct of this study and extension initiative. The authors also extend their appreciation to the Information Technology Department, the Extension Services and Community Engagement Unit, and Bukidnon State University for allowing and supporting the implementation of this community engagement program. Their institutional support, commitment to extension services, and dedication to empowering indigenous communities significantly contributed to the successful completion of this research.

## Generative AI Statement

During the preparation of this manuscript, the author(s) used ChatGPT by OpenAI to refine language, improve readability, and enhance academic writing structure. Following this process, the author(s) critically reviewed, edited, and validated the output to ensure accuracy and scientific integrity. The author(s) maintain full accountability for the final content. AI was used solely as a supportive tool and is not credited with authorship. No original data analysis or interpretations were performed by the AI without human oversight.